\documentclass[conference]{IEEEtran}
\IEEEoverridecommandlockouts

\usepackage{cite}
\usepackage{amsmath,amssymb,amsfonts}
\usepackage{algorithmic}
\usepackage{graphicx}
\usepackage{textcomp}
\usepackage{xcolor}
\usepackage{hyperref}
\def\BibTeX{{\rm B\kern-.05em{\sc i\kern-.025em b}\kern-.08em
    T\kern-.1667em\lower.7ex\hbox{E}\kern-.125emX}}
\begin{document}

\title{Systematic Evaluation of Time-Frequency Features for Binaural Sound Source Localization\\
\thanks{This publication has emanated from research conducted with the financial support of Taighde Éireann – Research Ireland under Grant numbers 12/RC/2289\_P2 and 13/RC/2077\_P2 and via a gift to support this research from Google. For the purpose of Open Access, the author has applied a CC BY public copyright license to any Author Accepted Manuscript version arising from this submission.}
}

\author{\IEEEauthorblockN{Davoud Shariat Panah$^1$, Alessandro Ragano$^1$, Dan Barry$^1$, Jan Skoglund$^2$, Andrew Hines$^1$}
\IEEEauthorblockA{
\textit{$^1$School of Computer Science, University College Dublin, Dublin, Ireland}\\
\textit{$^2$Google LLC, San Francisco, USA}\\
\{davoud.shariatpanah,alessandro.ragano\}@ucd.ie, danbarry@duck.com, jks@google.com, andrew.hines@ucd.ie
}
\and

}

\maketitle

\begin{abstract}
This study presents a systematic evaluation of time-frequency feature design for binaural sound source localization (SSL), focusing on how feature selection influences model performance across diverse conditions. We investigate the performance of a convolutional neural network (CNN) model using various combinations of amplitude-based features (magnitude spectrogram, interaural level difference - ILD) and phase-based features (phase spectrogram, interaural phase difference - IPD). Evaluations on in-domain and out-of-domain data with mismatched head-related transfer functions (HRTFs) reveal that carefully chosen feature combinations often outperform increases in model complexity. While two-feature sets such as ILD + IPD are sufficient for in-domain SSL, generalization to diverse content requires richer inputs combining channel spectrograms with both ILD and IPD. Using the optimal feature sets, our low-complexity CNN model achieves competitive performance. Our findings underscore the importance of feature design in binaural SSL and provide practical guidance for both domain-specific and general-purpose localization.
\end{abstract}


\section{Introduction}
\label{sec:intro}
Binaural audio captures spatial hearing cues by recording or synthesizing sound with two channels to replicate the auditory information humans use for three-dimensional localization. Binaural SSL leverages interaural cues such as time and level differences, along with spectral shaping from the outer ear, to determine the spatial position of sounds in three dimensions~\cite{blauert1997spatial}. Accurate source localization is crucial for natural hearing, enabling applications in virtual reality \cite{rychtarikova2009binaural}, auditory scene analysis \cite{ma2018robust}, and humanoid robots \cite{keyrouz2014advanced}.

SSL has seen significant advances in recent years, with deep learning models achieving state-of-the-art accuracy and robustness in complex acoustic environments. A variety of architectures have been successfully applied to the SSL problem, among which CNNs \cite{goli2023deep, vargas2021improved} and convolutional recurrent neural networks (CRNNs) \cite{grumiaux2021improved,nguyen2020sequence} are among the most widely used. More recently, a variety of Transformer models have been applied to the binaural SSL problem. For instance, Phokhinanan et al. \cite{phokhinanan2023binaural} proposed an adapted vision Transformer (ViT) model for binaural SSL in noisy conditions, using ILD and IPD as input feature representations. In another study \cite{phokhinanan2024auditory}, they introduced a ViT model with a spectral attention mechanism for binaural speech localization designed to handle mismatches in HRTFs between training and testing data. This model leveraged ILD, IPD, and magnitude spectrograms as input features. Kuang et al. \cite{kuang2025bast} proposed a Transformer model based on Mamba ViT \cite{hatamizadeh2025mambavision} for binaural SSL in both anechoic and reverberant conditions. This work used magnitude spectrograms as input features.

As the above examples demonstrate, previous work has employed a variety of feature representations with binaural audio data. The choice of features directly influences how effectively a binaural SSL model can extract spatial cues, remain robust in noisy and reverberant conditions, and generalize across environments \cite{bovbjerg2025learning}. Therefore, thoughtful feature design is often as important as the choice of model architecture. However, research in this area remains very limited, and to our knowledge, only one work has examined how the choice of input features affects SSL performance. Krause et al. \cite{krause2021feature} compared amplitude-based, phase-based, and covariance matrix-based features in terms of sound event detection and localization performance when used with a CRNN architecture. However, the data used in this study were collected with a tetrahedral microphone array, and therefore, the findings may not generalize to binaural data. Moreover, the study restricted feature combinations to pairs and did not explore larger sets that could potentially result in more effective representations and improved performance.

To address this gap, this study systematically investigates which feature set is more effective for binaural SSL using controlled experiments and cross-condition evaluation. Specifically, we present a comparative analysis of amplitude- and phase-based features, evaluating them individually and in combinations of up to four. Given that CNNs are among the most commonly used architectures in the field, we use these features as inputs to a CNN model. To assess the robustness of the features to content variability, we evaluate the models on both in-domain and out-of-domain content. Furthermore, we employ different sets of HRTFs from different subjects for training and testing to examine generalization across listening conditions. Finally, by making our source code and datasets publicly available, we provide the community with a benchmark resource that can accelerate future research, facilitate reproducibility, and support fair comparison of new methods.

\section{EXPERIMENTAL SETUP}
\label{sec:setup}
To investigate the impact of feature choice on SSL performance, we train multiple models using various feature combinations and systematically compare their performance on in-domain and out-of-domain data. Below, we detail the datasets, feature representations, model architecture, model training, and evaluation.

\subsection{Datasets}
\label{sec:datasets}  
We used the Binamix library \cite{barry2025binamix} to synthesize datasets for model training and evaluation. This library enables binaural mixing using the SADIE II database \cite{Armstrong2018Perceptual}, which provides head-related impulse responses (HRIRs) and binaural room impulse responses (BRIRs) across 20 subjects. To handle angles where measured impulse responses are unavailable, Binamix applies a modified Delaunay triangulation for interpolation.

The training set was generated by rendering randomly selected samples from the TSP speech dataset \cite{kabal2002tsp} with HRIRs from subjects D1, H3--H6, and H11--H13 in the SADIE database, covering the full sphere at 5° increments in both azimuth and elevation. This set contains 21,312 recordings. The validation set was generated in the same way using subjects H3 and H4. This set includes 5,328 recordings.

We also use three test sets in our experiments. The first is a speech-only test set, TSP--SSL, generated using the same procedure but with subjects H7--H10. To avoid using the same content, we ensured that speakers from the TSP dataset did not overlap across the training, validation, and test sets. The TSP--SSL test set includes 10,656 recordings. The second test set is the localization sensitivity subset from the SynBAD dataset \cite{panah2025binaqual}, which contains a variety of natural (e.g., castanets) and synthetic (e.g., pink noise) content spatialized using HRIRs from subject D2 of the SADIE database. This set has a variable angular resolution that becomes coarser toward the periphery, aligned with the characteristics of human perceptual localization accuracy, which is highest in the frontal region. It includes 11,500 recordings, and we refer to it as SynBAD--Var. The third test set is derived from the same content as SynBAD--Var, but rendered at a uniform 5° resolution across both azimuth and elevation. This test set includes 16,560 recordings, and we refer to it as SynBAD--Fix. We use the TSP--SSL as an in-domain test set and both SynBAD sets as out-of-domain sets.

As mentioned above, the data in all sets cover the full azimuthal range (0° -- 360°). We map azimuth angles to the frontal plane (-90° -- 90°) by wrapping angles appropriately. All sets were originally sampled at 48 kHz but were resampled to 16 kHz for processing. The datasets are available online\footnote{\url{https://zenodo.org/records/20355674}}.

\subsection{Feature Representations}
\label{sec:features}
We use four core features: two amplitude-based (magnitude spectrogram and ILD) and two phase-based (phase spectrogram and IPD).

\textit{Magnitude spectrogram} represents the amplitude of the short-time Fourier transform (STFT) of a signal:
\begin{equation}
X(t, f) = \sum_{n=-\infty}^{\infty} x[n] \, w[n - tH] \, e^{-j 2 \pi f n / N},
\end{equation}
where $x[n]$ is the time-domain signal, $w[\cdot]$ is the analysis window, $H$ is the hop size, and $N$ is the FFT length. The magnitude spectrogram is then:
\begin{equation}
M(t,f) = |X(t,f)|.
\end{equation}

\textit{Phase spectrogram} represents the instantaneous phase of the STFT:
\begin{equation}
\Phi(t,f) = \arg\big(X(t,f)\big),
\end{equation}
where $\arg(\cdot)$ denotes the argument (phase angle) of the complex STFT coefficients.

\textit{ILD} captures the difference in level between the left and right ear signals in the frequency domain:
\begin{equation}
\text{ILD}(t,f) = 20 \log_{10} \frac{|X_L(t,f)|}{|X_R(t,f)|},
\end{equation}
where $X_L(t,f)$ and $X_R(t,f)$ are the STFTs of the left and right ear signals. According to the duplex theory \cite{wightman1992dominant}, ILD is most effective for high-frequency sounds, generally above 1.5 kHz, where the head casts an acoustic shadow, producing measurable amplitude differences between the ears. At low frequencies, the head shadow effect is negligible, and as a result, ILD provides little localization information. 

\textit{IPD} encodes the phase difference between the left and right ear signals:
\begin{equation}
\text{IPD}(t,f) = \Phi_L(t,f) - \Phi_R(t,f),
\end{equation}
where $\Phi_L(t,f) = \arg(X_L(t,f))$ and $\Phi_R(t,f) = \arg(X_R(t,f))$. The difference is wrapped to $[-\pi, \pi]$. IPD is a reliable cue for low-frequency sounds, typically below 1.5 kHz. At higher frequencies, phase ambiguity arises, reducing the usefulness of IPD for localization \cite{wightman1992dominant}.

To compute the STFT for the above features, a window length of 25 ms was used, corresponding to 400 samples at a 16 kHz sampling rate. Also, a hop length of 10 ms (160 samples) was applied between consecutive frames. These features are combined in different ways to form multi-channel inputs to a CNN model described in the next section. 

\begin{table*}[ht]
\small
\setlength{\abovecaptionskip}{1pt}
\setlength{\belowcaptionskip}{1pt}
\centering
\caption{Comparison of feature performance (MAE $\downarrow$) across two test sets with the CNN model. Abbreviations: Mag – Magnitude spectrogram, Phase – Phase spectrogram, L – Left, R – Right, El – Elevation, AS – All Sources, NS – Natural Sources.}
\renewcommand{\arraystretch}{1}
\setlength{\tabcolsep}{6pt}
\begin{tabular}{|l|c|l|l|l|l|l|}
\hline
& & \multicolumn{2}{c|}{\textbf{TSP--SSL test set}} 
& \multicolumn{3}{c|}{\textbf{SynBAD--Var test set}} \\ \hline
\textbf{Feature set} & \textbf{\# Features} 
& All El & El = 0° & AS, All El & AS, El = 0° & NS, El = 0° \\ \hline
ILD & 1 & 11.5° & 7.1° & 22.7° & 23.2° & 14.3° \\ 
IPD & 1 & 12.1° & 6.7° & 21.8° & 21.5° & 10.2° \\ 
Mag L/R & 1 & 13.3° & 6.3° & 31.7° & 29.9° & 12.4° \\ 
Phase L/R & 1 & 13.6° & 8.4° & 43.1° & 39.0° & 8.1° \\ 
ILD, IPD & 2 & 10.1° & \textbf{4.5°} & 14.2° & 10.7° & 6.0° \\ 
Mag L/R, ILD & 2 & 10.7° & 4.8° & 18.9° & 16.3° & 9.6° \\ 
Mag L/R, IPD & 2 & 10.1° & \textbf{4.5°} & 17.1° & 13.8° & 9.6° \\ 
Phase L/R, ILD & 2 & 10.4° & 4.9° & 18.8° & 17.9° & 7.2° \\ 
Phase L/R, IPD & 2 & 11.5° & 5.9° & 17.9° & 11.8° & 8.4° \\ 
Mag L/R, Phase L/R & 2 & 10.7° & 5.0° & 23.8° & 21.9° & 13.5° \\ 
Mag L/R, ILD, IPD & 3 & \textbf{10.0°} & 4.6° & 15.9° & 11.6° & 6.5° \\ 
Phase L/R, ILD, IPD & 3 & 10.1° & 4.6° & \textbf{12.7°} & \textbf{8.4°} & \textbf{4.0°} \\ 
Mag L/R, Phase L/R, ILD, IPD & 4 & \textbf{10.0°} & \textbf{4.5°} & 15.0° & 10.3° & 5.8° \\ \hline
\end{tabular}
\label{tab:features_comparison}
\end{table*}

\subsection{Model Architecture}
\label{sec:model_arch}
We use a CNN architecture for binural SSL. The network is designed to process multi-channel input features and extract hierarchical spatial representations. 

The architecture consists of three convolutional layers with kernel size $3\times3$, stride 1, and padding 1. The first, second, and third convolutional layers have 32, 64, and 128 output channels, respectively. Each convolutional layer is followed by a ReLU activation and a $2\times2$ max-pooling operation. After the final convolutional block, a global average pooling layer is applied, producing a 128-dimensional feature vector. A fully connected layer with 128 units and ReLU activation produces the final features. A dropout layer (rate 0.3) is applied to improve generalization. This architecture efficiently extracts spatial cues from multi-channel inputs while maintaining a relatively lightweight design. This simple design allows us to ensure our results reflect the quality of the features, not the complexity of the model.

\subsection{Model Training}
\label{sec:model_train}

For training the model to predict azimuth, we use a loss function that explicitly accounts for the circular nature of angles. Let $\theta$ denote the ground-truth azimuth and $\hat{\theta}$ denote the predicted azimuth, both expressed in degrees and converted to radians. The shortest angular difference is computed as:

\begin{equation}
\Delta \theta_{\text{az}} = \text{atan2}\Big(\sin(\theta - \hat{\theta}), \cos(\theta - \hat{\theta})\Big),
\end{equation}

where $\Delta \theta_{\text{az}} \in [-\pi, \pi]$.  
The final azimuth loss is obtained by applying the mean squared error (MSE) to this angular difference across all $N$ samples:

\begin{equation}
\mathcal{L}_{\text{az}} = \frac{1}{N} \sum_{i=0}^{N-1} (\Delta \theta_{\text{az},i})^2.
\end{equation}

The network is trained using the Adam optimizer for a maximum of 1000 epochs using a learning rate of 0.001. Training is terminated if the validation loss does not improve for 20 consecutive epochs. The model was implemented using the PyTorch 2.8 library. The source code is available online\footnote{\url{https://github.com/dspanah/Binaural_SSL}}.

\subsection{Evaluation}
\label{sec:evaluation}
To evaluate the localization performance of the models, we use the mean angular error (MAE) metric, which quantifies the average angular error between predicted and ground-truth azimuths. MAE in degrees is computed as:


\begin{equation}
\text{MAE} = \frac{180}{\pi} \cdot \frac{1}{N} \sum_{i=0}^{N-1} \big| \Delta \theta_{\text{az},i} \big|.
\end{equation}

Smaller values of $\text{MAE}$ indicate more accurate localization predictions.

We compare the performance of our best-performing models with three recent binaural SSL models as baselines: FAViT \cite{phokhinanan2023binaural}, AMViT \cite{phokhinanan2024auditory}, and BAST-MAMBA \cite{kuang2025bast}.

\begin{figure*}[]
  \centering
  \includegraphics[width=0.99\linewidth]{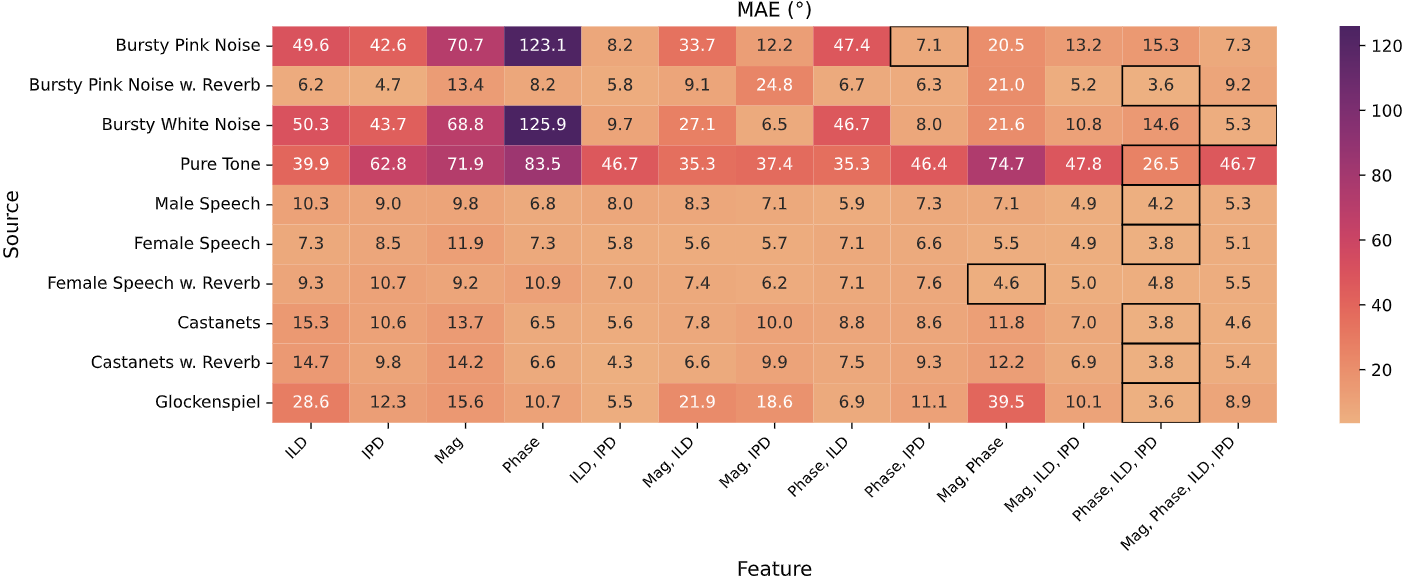}
  \caption{MAE $ \downarrow $
of the features across sources on the SynBAD--Var test set for the CNN model - Elevation = 0°.}

  \label{fig:synbad_var}
\end{figure*}

\section{RESULTS and Discussion}
\label{sec:results}

\subsection{Impact of Feature Representations}
\label{sec:results_ifr}
Table \ref{tab:features_comparison} summarizes the MAE of all feature sets on the TSP--SSL (in-domain) and SynBAD--Var (out-of-domain) test sets. On the TSP--SSL, single features including ILD, IPD, Mag L/R, and Phase L/R show the highest localization errors. Combining these features reduces the error. For example, the two-feature ILD + IPD representation achieves an MAE of 4.5° (El = 0), highlighting the complementary role of low- and high-frequency cues for robust localization. Similarly, combining magnitude spectrograms (Mag L/R) with phase spectrograms (Phase L/R) outperforms the individual features, though it is less effective than ILD + IPD. Among the best-performing configurations, several feature sets achieve comparable performance, indicating that compact two-feature inputs such as ILD + IPD can be as effective as four-feature inputs. This result can be attributed to the fact that the TSP--SSL set contains only speech, meaning that training and test domains are matched. We also observe that the three- and four-feature sets do not provide a substantial improvement over the two-feature ILD + IPD combination. This suggests that, for this in-domain task, diminishing returns are reached quickly, and adding more features does not necessarily result in better performance.

The SynBAD--Var test set is significantly more challenging than the TSP--SSL, as it includes a variety of synthetic and natural sounds (e.g., pink noise, castanets, pure tones, speech with reverb) that differ from the training data. Similar to the TSP--SSL, on this test set, single features such as Phase L/R or Mag L/R yield the highest localization errors. As before, combining ILD and IPD provides a substantial improvement, reducing the MAE to 10.7° (all sources, elevation = 0°). Unlike the speech-only test set, adding more features leads to considerable performance gains on the SynBAD--Var test set. The best-performing configuration is the three-feature combination Phase L/R + ILD + IPD, which achieves the lowest errors across all three scenarios: (i) all sources, all elevations; (ii) all sources at 0° elevation; and (iii) natural sources at 0° elevation. These results indicate that to generalize across diverse and out-of-domain content, a richer feature set is essential. In particular, the combination of channel phase spectrograms with both ILD and IPD consistently delivers the most accurate localization, demonstrating the value of pairing raw signal representations with explicit interaural cues.

Fig. \ref{fig:synbad_var} illustrates the MAE of the features across different sources (content) on the SynBAD--Var test set (elevation = 0°). The first four sources are synthetic signals that lack the complex harmonic and temporal structure of natural sounds. These sources, particularly the anechoic broadband noise signals, act as a stress test for the feature representations and reveal their limits of generalization. The monaural features (magnitude and phase spectrograms) fail on these signals, with the phase spectrogram reaching an MAE of over 125° for white noise. This indicates that a model trained on structured speech cannot extract meaningful spatial information from raw spectrograms of unstructured signals. In contrast, the two-feature ILD + IPD achieves an MAE below 10° for the same signals, demonstrating that explicit binaural cues provide a considerably more robust representation. Interestingly, the Bursty Pink w. Reverb source is localized more accurately than its anechoic counterpart across almost all feature sets. This indicates that the early reflections introduced by reverberation provide additional temporal and spectral structure, enabling the model to improve localization on otherwise feature-poor signals. The Pure Tone source is a uniquely difficult case. As shown in Fig. \ref{fig:synbad_var}, almost every feature combination struggles, resulting in high errors. This is consistent with the principles of psychoacoustics, where a single sine wave can create ambiguous localization cues \cite{blauert1997spatial}.

Conversely, the model shows strong generalization across all natural sound sources, including transient sounds such as castanets and tonal sounds like the glockenspiel. A key observation is the consistent performance of the Phase L/R + ILD + IPD feature set, which combines channel phase spectrograms with binaural cues. This configuration achieves the lowest errors across speech, castanets, and glockenspiel sources. The strong generalization to castanets and glockenspiel is significant, as these sounds exhibit very different temporal characteristics from speech, with sharp, percussive onsets. Overall, these results indicate that robust localization across diverse real-world sounds is best achieved when the model is provided with both raw signal context (via spectrograms) and explicit interaural cues (ILD and IPD), allowing it to learn features that are not specific to a single sound category.

\subsection{Comparison With Previous Work}
\label{sec:results_cwpw}
Table \ref{tab:prev_work} compares our model, using its best-performing feature combination for each test set, against three recent Transformer-based models: FAViT \cite{phokhinanan2023binaural}, AMViT \cite{phokhinanan2024auditory}, and BAST-MAMBA \cite{kuang2025bast}. Out of these, the first two are classification models, while the third one is a regression model. To ensure a fair comparison with classification models, we also report the results on the SynBAD--Fix test set, which has a uniform resolution of 5°, similar to our training set.

As shown in Table \ref{tab:prev_work}, on the in-domain TSP--SSL test set, the FAViT model, which leverages ILD and IPD features, achieves the best performance. This indicates that for speech localization, the attention mechanism of a vision Transformer can be highly effective when provided with a representation of the core spatial cues. In contrast, the poor performance of BAST-MAMBA, which relies solely on magnitude spectrograms, suggests that learning interaural relationships directly from channel spectrograms is substantially more challenging, even for a high-complexity sequence model.

On the more challenging out-of-domain SynBAD test sets, our low-complexity CNN outperforms all baselines despite having far fewer parameters. By leveraging a rich combination of channel spectrograms and binaural cues, it achieves an MAE of 8.4°, making it the most robust model for diverse out-of-domain content. We should note that FAViT and AMViT frame SSL as a classification task with 5° azimuth bins, which inherently limits their output resolution to the bin width. In contrast, our model performs continuous-angle regression, allowing us to pursue angular precision finer than 5° when the signal and conditions permit. Overall, these results highlight that robust feature design is more effective for binaural SSL than simply increasing model complexity. In fact, relying solely on a complex architecture proves insufficient, as evidenced by the weaker performance of larger Transformer-based models.

\setlength{\tabcolsep}{4pt}
\begin{table}[]
\centering
\caption{Comparing the \text{MAE} $\downarrow$ $\pm$ 95\% CI
 of the best-performing feature on each test set with previous work - Elevation = 0°.}
\begin{tabular}{lllll}
\hline
 & TSP--SSL & SynBAD--Var & SynBAD--Fix & \# Params \\
\hline
FAViT      & \textbf{2.7° $\pm$ 0.4°} & 27.5° $\pm$ 3.8 & 26.1° $\pm$ 3.1° & 0.6 M \\
AMViT      & 3.7° $\pm$ 0.8° & 15.8° $\pm$ 2.3° & 14.2° $\pm$ 1.8° & 1.3 M \\
BAST-MAMBA & 10.7° $\pm$ 1.4° & 11.0° $\pm$ 0.9° & 13.2° $\pm$ 0.9° & 26.0 M \\
CNN (ours)        & 4.5° $\pm$ 0.4° & \textbf{8.4° $\pm$ 1.0°} & \textbf{8.7° $\pm$ 0.9°} & \textbf{0.1 M} \\
\hline
\end{tabular}
\label{tab:prev_work}
\end{table}

\section{Conclusions}
\label{sec:conclusion}
This study presented a systematic evaluation of feature design for binural SSL, emphasizing the significant role of careful feature selection in achieving robust and accurate localization across varied conditions. Through controlled experiments, we demonstrated that the choice of input features often has a greater impact on performance than model complexity: well-chosen, complementary features can lead to greater gains than scaling up the model.

On in-domain speech data, two-feature sets such as ILD + IPD achieve localization accuracy comparable with larger combinations, indicating that minimal yet complementary cues suffice when the train and test domains align. However, under challenging out-of-domain conditions, richer feature sets are essential. The combination of Phase L/R, ILD, and IPD consistently delivers the best generalization, especially for structurally diverse signals such as high-frequency instrument sounds and reverberant noise.

Having established the strength of these feature sets, future work will explore their integration into more complex architectures to assess whether they can further enhance overall SSL performance. Additionally, we will evaluate the effectiveness of these features in more complex scenarios, such as multi-source localization.



\bibliographystyle{ieeetr}
\bibliography{refs.bib}


\end{document}